\documentclass[final,journal,letterpaper]{IEEEtran}

 \usepackage{varioref} 
 \usepackage{wrapfig} 
 \usepackage{threeparttable} 
 \usepackage{dcolumn} 
  \newcolumntype{d}{D{.}{.}{-1}}
 \usepackage{nomencl} 
  \makeglossary
 \usepackage{subfigmat} 
 \usepackage{fancyvrb} 
  \fvset{fontsize=\footnotesize,xleftmargin=2em}
 \usepackage{lettrine} 
 \usepackage[hidelinks]{hyperref}

 \usepackage{epsfig}
 \usepackage{epstopdf}
 \usepackage{subfigure}
 \usepackage{latexsym}
 \usepackage{amsmath}
 \usepackage{amsfonts}
 \usepackage{amssymb}
 \usepackage{mathrsfs}
 \usepackage{cases}
 \usepackage{array}

 \usepackage{url}
 \usepackage[noadjust]{cite}
 \usepackage{graphicx}
 \usepackage{psfrag}
 \usepackage{color}

\newtheorem{remark}{Remark}

\newtheorem{case}{Case}

\newcommand{\mb}{\mbox}
\newcommand{\bm}{\boldmath}
\def\b#1{\mb{\bm$#1$}}
\newfont{\Bb}{msbm10 scaled\magstep1}

\newcommand{\diag}{\mathop{\mathrm{diag}}}

\begin{document}

\title{Decentralized Robust Control for Damping Inter-area Oscillations in Power Systems}


\author{Jianming Lian, Shaobu Wang, Ruisheng Diao and Zhenyu Huang%
\thanks{J. Lian, S. Wang, R. Diao and Z. Huang are with Pacific Northwest National Laboratory, Richland, WA 99352, USA (e-mail: \{jianming.lian, shaobu.wang, ruisheng.diao, zhengyu.huang\}@pnnl.gov).}}

\maketitle

\begin{abstract}
As power systems become more and more interconnected, the inter-area oscillations has become a serious factor limiting large power transfer among different areas. Underdamped (Undamped) inter-area oscillations may cause system breakup and even lead to large-scale blackout. Traditional damping controllers include Power System Stabilizer~(PSS) and Flexible AC Transmission System~(FACTS) controller, which adds additional damping to the inter-area oscillation modes by affecting the real power in an indirect manner. However, the effectiveness of these controllers is restricted to the neighborhood of a prescribed set of operating conditions. In this paper, decentralized robust controllers are developed to improve the damping ratios of the inter-area oscillation modes by directly affecting the real power through the turbine governing system. The proposed control strategy requires only local signals and is robust to the variations in operation condition and system topology. The effectiveness of the proposed robust controllers is illustrated by detailed case studies on two different test systems.
\end{abstract}


\section{Introduction}\label{sec:Intro}
The inter-area oscillations of low frequency are inherent phenomena between synchronous generators that are interconnected by transmission systems. It has been pointed out in~\cite{Klein91} that these oscillations often become poorly damped with heavy power transfer between different areas over weak connections of long distance. Because sustained oscillations could cause system breakup and even lead to large-scale blackout, it is extremely important to maintain the stability of these oscillations for the system security. However, environmental constraints and economic pressures often push power systems close to their operational limits~\cite{Bertsch05}, which substantially reduces the stability margin of the normal operation. One of the prominent events caused by undamped oscillations is the notable breakup and blackout in the western North American Power System on August 10, 1996~\cite{Kosterev99}.
To ensure the secure system operation, the amount of power transfer on tie-lines between different areas are often limited due to the inter-area oscillations that are poorly damped. Hence, new control strategies that can improve the damping of these inter-area oscillations are needed to increase the transmission capacity for a better utilization of the existing transmission network.

The well-known damping controller is the power system stabilizer~(PSS)~\cite{Kundur94}. It adds a supplementary control signal into the generator excitation system through the automatic voltage regulator~(AVR) to apply an additional electric torque on the generator rotor. However, the classical design of the PSS, which relies on the linearization of a single-machine infinite-bus~(SMIB) model, has two major shortcomings. First, the SMIB model is a reduced-order model by neglecting those important dynamics associated with network interconnections. It cannot adequately describe the inter-area oscillation modes due to network interconnections. Hence, the classical PSS is effective in damping local oscillation, but it may not provide enough damping to the inter-area oscillations unless carefully tuned. Furthermore, the local controller design is independent of one another without any proper coordination among them. Second, the linearization based controller design greatly restricts the validity of the controller to the neighborhood of the operating point under consideration. When the system loading or network topology drastically changes during the normal system operation, the resulting PSS may no longer yield satisfactory damping effect.

Another approach of damping the inter-area oscillations is to introduce various supplementary modulation controllers to flexible AC transmission system~(FACTS) devices (see, for example,~\cite{Lerch91,Yang98,El-Moursi10,Zarghami10}). Although FACTS controllers can achieve satisfactory damping effect for inter-area oscillations~\cite{Mithulananthan03}, it is not cost-effective to use them for the sole purpose of damping control. On the other hand, FACTS controllers share the same shortcoming with the classical PSS due to the linearization based controller design. Furthermore, the interactions between FACTS controllers and the PSSs could even potentially degrade the damping without proper coordination~\cite{Gibbard00}.

Recently, there have been a lot of efforts into the development of coordinated and robust PSSs for power systems~\cite{Abdel-Magid00,Zhang12,Rao00,Boukarim00}. In~\cite{Abdel-Magid00}, robust tuning based on eigenvalue analysis was proposed to select the parameters of PSSs over a prescribed set of operating conditions. In~\cite{Zhang12}, a novel method based on modal decomposition was proposed to eliminate the interactions among different modes for tuning PSSs. In~\cite{Rao00} and~\cite{Boukarim00}, robust PSSs were designed by utilizing robust control techniques such as $\mu$-synthesis and linear matrix inequality~(LMI). In addition to robust PSSs, robust excitation control strategies have also been developed for damping control and the resulting controllers have different structure from that of the classical PSS. In~\cite{Ramos04}, a novel methodology was proposed to design robust excitation controller based on the polytopic model for improved robustness over varying operating conditions. Although these newly developed robust damping controllers are more robust than the classical PSS, their robustness is still limited to a finite number of operating conditions.


There have been also persistent efforts on determining appropriate global signals as the control input of classical PSSs and FACTS controllers as in~\cite{Aboul-Ela96} and~\cite{Farsangi04}. Because the feedback of global signals actually establishes the correlation among independent local controllers, it is more effective than the feedback of local signals only. These efforts have been particularly facilitated by the technology of wide area measurement system~(WAMS) enabled by the development of Phasor Measurement Unit~(PMU). Many WAMS-based damping controllers have been developed (see, for example,~\cite{Kamwa01,Zhang08,Li12,Trudnowski12}). However, the robustness of WAMS-based damping controllers to the variations in operating condition and system topology still requires further investigated. Moreover, the destabilizing factors of communication network such as time-varying delay and random packet dropout has to be explicitly taken into account for the reliability purpose~\cite{Trudnowski08report}, which greatly complicates the corresponding controller design.

In most cases, the inter-area oscillations are usually caused by transferring real power between interconnected areas. Thus, it is a straightforward way to control the real power in the system for the oscillation mitigation. However, the majority of existing damping controllers introduce additional damping to the inter-area oscillation modes through either generation excitation system (e.g., PSS) or transmission line (e.g., FACTS), which actually affects the real power in an indirect manner. Some of the work that {\it directly} affects the real power to damp the inter-area oscillations has been reported in~\cite{Pal00,Neely13,Huang10}. In~\cite{Pal00}, robust damping controllers were designed for superconducting magnetic energy storages to enhance the damping of multiple inter-area modes by injecting real power into the system. It is shown that real power modulation can be an effective way to damp power flow oscillation, although the resulting controller therein is only robust to the prescribed set of operating conditions. In~\cite{Neely13}, two damping control systems using ultracapacitor based energy storages for real power injection were designed for areas expected to oscillate against one another. Because the frequency measurement of one area is required by the controller in the other area, the communication effects could greatly degrade the expected controller performance. In~\cite{Huang10}, a Modal Analysis for Grid Operations~(MANGO) procedure was established to enable grid operators to immediately mitigate the inter-area oscillations by adjusting operating conditions, where the detection of low damping is achieved by the modal analysis with real-time phasor measurements. The MANGO procedure can enact as part of a remedial action scheme. However, the interactions among different oscillation modes make it challenging to determine universal dispatch patterns that are effective for all the inter-area oscillation modes.

In this paper, the decentralized robust control strategy developed in~\cite{Marinovici13} is adopted herein to improve the damping ratios of the inter-area oscillation modes. In~\cite{Marinovici13}, a distributed hierarchical control architecture was proposed for large-scale power systems to improve the transient stability and frequency restoration. However, the capability of this control strategy in enhancing the small signal stability of multi-area power systems has not been analyzed yet. Hence, it is the main focus of this paper to show the effectiveness of this control strategy in mitigating the inter-area oscillations. This robust controller introduces an auxiliary control signal into the turbine governing system through the governor to apply an additional mechanical torque on the generator rotor. This principle is similar to that of the PSS, whereas the PSS provides an additional electric torque on the generator rotor through the generator excitation system. The proposed control strategy has several {\it appealing advantages} ascribed to the underlying coordinated controller design. {\it First}, it is robust to the variations in operating condition and system topology. This is because the corresponding controller design does not require system linearization over a specific operating condition, nor does it depend on a specific network topology either. Such robustness is extremely crucial for the secure operation of power systems, which constantly experience changes in operating conditions and transmission networks. {\it Second}, it has no destructive interference with existing PSSs in the system regarding damping control, which eliminates the need of proper coordination. This is because the turbine governing system is weakly coupled with the excitation system. {\it Third}, it is decentralized requiring only local signals that are easily obtained for controller implementation, which greatly improves the reliability of the controller by eliminating the use of communication. {\it Lastly}, it has simple controller design of low complexity and does not involve additional costly equipments such as power electronics and energy storages, which make it easy for practical implementation. Detailed case studies are performed on both small and medium-sized test systems to illustrate the effectiveness and robustness of the proposed robust controllers in improving the damping ratios of the inter-area oscillation modes.




The remainder of this paper is organized as follows. In Section~\ref{sec:modeling}, the dynamic model of multi-area power systems is given. In Section~\ref{sec:DRC}, the design of decentralized robust controllers is described with the discussion of practical implementation. Then detailed case studies are presented in Section~\ref{sec:case}. Conclusions are found in Section~\ref{sec:conclusion}.

\section{System Modeling}\label{sec:modeling}

\begin{figure}
\centering
\subfigure[Rotor dynamics]{\includegraphics[width=0.375\textwidth]{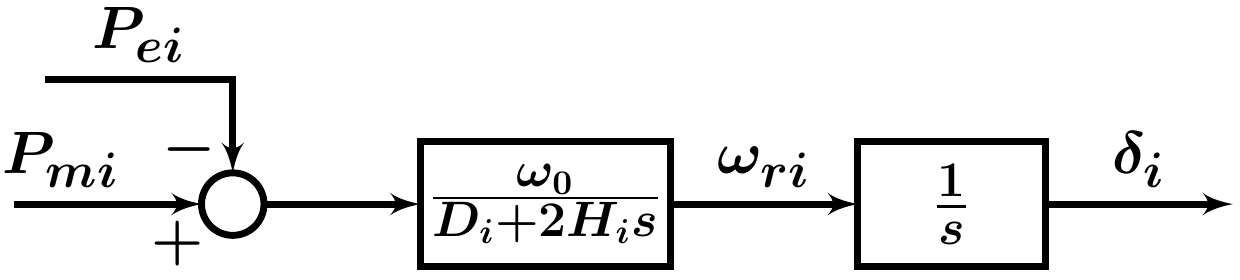}}
\subfigure[Governor control]{\includegraphics[width=0.38\textwidth]{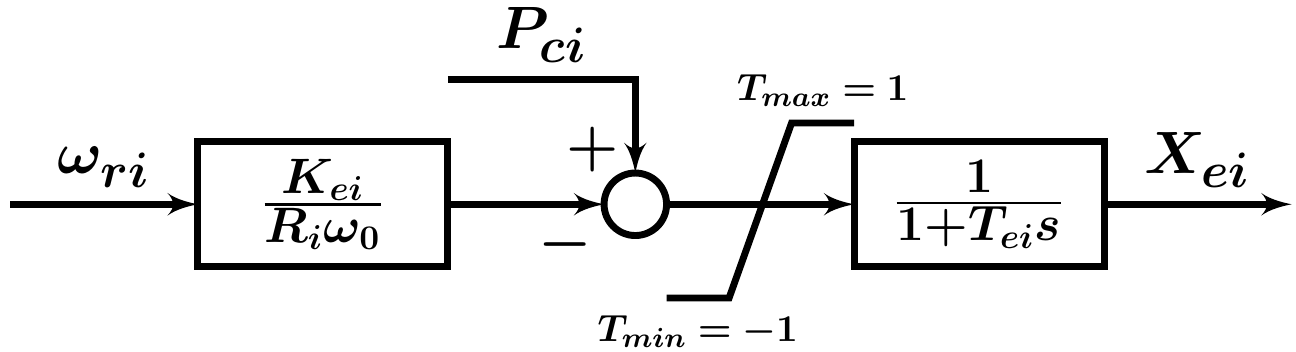}}
\subfigure[Turbine dynamics]{\includegraphics[width=0.35\textwidth]{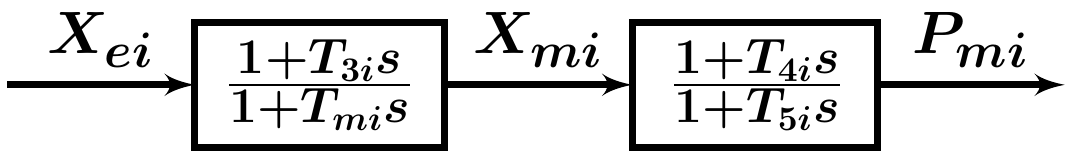}}
\caption{Generator dynamics with steam valve governor control~\cite{Marinovici13}.}
\label{fig:generator}
\end{figure}

The multi-area power systems usually consist of a number of generators that are interconnected through a transmission network. In this paper, robust controllers are proposed for generators with steam valve governors. Although only steam valve governors are considered herein, the proposed control strategy can be extended to other types of governors such as hydraulic governors. However, the resulting controller structures could vary due to different turbine governing system dynamics to be considered, which is left for future research. As illustrated in Fig.~\ref{fig:generator},
the generator dynamics with steam valve governor can be described by the following differential equations:
\begin{itemize}
\item[] \emph{Rotor Dynamics:}
    \begin{align}
    \dot{\delta}_i &= \omega_{ri}=\omega_i-\omega_o,\label{eqn:deltai}\\
    \dot{\omega}_{ri} &= -\frac{D_i}{2H_i}\omega_{ri}+\frac{\omega_o}{2H_i}
    \left(P_{mi}-P_{ei}\right);\label{eqn:omegai}
    \end{align}

\item[] \emph{Governor Control:}
    \begin{equation}\label{eqn:xei}
    \dot{X}_{ei}=-\frac{K_{ei}}{T_{ei}R_i\omega_o}\omega_{ri}-
    \frac{1}{T_{ei}}X_{ei}+\frac{1}{T_{ei}}P_{ci},
    \end{equation}
    \begin{equation*}
    0\le X_{ei}\le 1;
    \end{equation*}

\item[] \emph{Turbine Dynamics:}
	\begin{align}
    \dot{X}_{mi}=&-\frac{K_{ei}T_{3i}}{T_{mi}T_{ei}R_{i}\omega_{0}}\omega_{ri}-\frac{1}{T_{mi}}X_{mi}
    \nonumber \\
    &+\frac{1}{T_{mi}}\left(1-\frac{T_{3i}}{T_{ei}}\right)X_{ei}+\frac{T_{3i}}{T_{mi}T_{ei}}P_{ci},
    \label{eqn:xmi} \\
    \dot{P}_{mi}=&-\frac{K_{ei}T_{3i}T_{4i}}{T_{mi}T_{ei}T_{5i}R_{i}\omega_{0}}\omega_{ri}-\frac{1}{T_{5i}}P_{mi}
    \nonumber \\
    &+\frac{1}{T_{5i}}\left(1-\frac{T_{4i}}{T_{mi}}\right)X_{mi}+\frac{T_{3i}T_{4i}}{T_{mi}T_{ei}T_{5i}}P_{ci}\nonumber \\
    &+\frac{T_{4i}}{T_{mi}T_{5i}}\left(1-\frac{T_{3i}}{T_{ei}}\right )X_{ei}; \label{eqn:pmi}
    \end{align}
\end{itemize}
In the above dynamics, $\delta_i$ is the generator rotor angle (rad), $\omega_i$ is the generator rotor speed (rad/sec), $\omega_{ri}$ is the relative rotor speed (rad/sec), $\omega_0$ is the nominal rotor speed (rad/sec), $P_{mi}$ is the mechanical power (p.u.) provided by turbine, $P_{ei}$ is the electrical power (p.u.) resulted from network interconnections, $X_{ei}$ is the valve opening (p.u.), and $P_{ci}$ is power control input (p.u.). For the normal operation of power systems, $P_{ci}=P_{ci}^{ref}$, which is the prescribed generator reference power received from economical dispatch.

When power systems do not have enough damping, any variations in $P_{ei}$ resulted from the changes in either loading condition or network topology can cause poorly damped oscillations of low frequency. In this work, the proposed decentralized robust controllers will modulate the mechanical power $P_{mi}$ to introduce additional damping to the poorly damped oscillation modes. In the following section, the controller design will be briefly presented. The interested readers are referred to~\cite{Marinovici13} for more details.

\section{Decentralized Robust Control}\label{sec:DRC}
To proceed, suppose there are $N$ generators in the system. Let $\b x_i=[\delta_i\ \omega_{ri}\ P_{mi}\ X_{mi}\ X_{ei}]^\top$ and $\b x=[\b x_1\ \ldots\ \b x_N]^\top$. Then the dynamics of the $i$-th generator can be represented by the following state space model,
\begin{equation}\label{eqn:sysx1}
\dot{\b x}_i=\b A_i\b x_i+\b B_i P_{ci}+\b G_iP_{ei},
\end{equation}
where $\b A_{i}$, $\b B_{i}$ and $\b G_{i}$ can be derived from~\eqref{eqn:deltai} through~\eqref{eqn:pmi} and are given by\renewcommand{\arraystretch}{1.5}\setlength{\arraycolsep}{1.75pt}
\begin{equation*}
\b A_{i}=\left[\begin{array}{ccccc}
0 & 1 & 0 & 0 & 0\\
0 & -\frac{D_{i}}{2 H_{i}} & \frac{\omega_{0}}{2 H_{i}} & 0 & 0\\
0 & -\frac{K_{ei} T_{3i} T_{4i}}{T_{mi} T_{ei} T_{5i} R_{i} \omega_{0}} & -\frac{1}{T_{5i}} & \frac{T_{mi} - T_{4i}}{T_{5i} T_{mi}} & \frac{T_{4i} \left ( T_{ei} - T_{3i} \right )}{T_{mi} T_{5i} T_{ei}}\\
0 & -\frac{K_{ei} T_{3i}}{T_{mi} T_{ei} R_{i} \omega_{0}} & 0 & -\frac{1}{T_{mi}} & \frac{T_{ei} - T_{3i}}{T_{mi} T_{ei}}\\
0 & -\frac{K_{ei}}{T_{ei} R_{i} \omega_{0}} & 0 & 0 & -\frac{1}{T_{ei}}
\end{array}\right],
\end{equation*}\setlength{\arraycolsep}{5.0pt}
\begin{equation*}
\b B_{i}=\begin{bmatrix}
0 & 0 & \frac{T_{3i} T_{4i}}{T_{mi} T_{ei} T_{5i}} & \frac{T_{3i}}{T_{mi} T_{ei}} & \frac{1}{T_{ei}}
\end{bmatrix}^\top,
\end{equation*}
\begin{equation*}
\b G_{i}=\begin{bmatrix}
0 & -\frac{\omega_{0}}{2 H_{i}} & 0 & 0 & 0
\end{bmatrix}^\top.
\end{equation*}\renewcommand{\arraystretch}{1.0}For each generator, the original power control input $P_{ci}^{ref}$ will be augmented with a local decentralized control $u_i$ so that
\begin{equation*}
P_{ci}=P_{ci}^{ref}+u_i.
\end{equation*}
This additional control input $u_i$ is defined as
\begin{equation}\label{eqn:controller}
u_{i}=\b k_i\left(\b x_i-\b x_i^d\right),
\end{equation}
where $\b x_i^d=[\delta_i^d\ \omega_{ri}^d\ P_{mi}^d\ X_{mi}^{d}\ X_{ei}^d]^\top$ is an arbitrary operating point, and $\b k_i$ is the linear feedback gain matrix. In practice, $\b x_i^d$ can be always chosen as the current equilibrium state of the system before any disturbances.
\begin{remark}
The state space model~\eqref{eqn:sysx1} is derived without linearization since all the nonlinearities resulted from network interconnections are encompassed by the unknown input $P_{ei}$. In the following controller design, the variations in $P_{ei}$ will be treated as external disturbances to individual generators and then rejected by the proposed controllers. The determination of controller gain $k_i$ in~\eqref{eqn:controller} is independent of any specific operating condition or network topology in order to guarantee the controller robustness. The operating point $\b x_i^d$ in~\eqref{eqn:controller} only serves as the reference from which the control input should be calculated. If the proposed controllers are activated during the normal operation, $\b x_i^d$ will be the current equilibrium operating point. If they are activated during the disrupted operation, $\b x_i^d$ will be the last equilibrium operating point before the disruption occurs. Once activated in the system, the proposed controllers will continue to work regardless of the operating condition and network topology.
\end{remark}

\subsection{Controller Design}
Let $\b x_i^e=[\delta_i^e\ \omega_{ri}^e\ P_{mi}^e\ X_{mi}^{e}\ X_{ei}^e]^\top$ denote the equilibrium state the system reaches after disturbances. It follows from~\eqref{eqn:sysx1} and~\eqref{eqn:controller} that $\b x_i^e$ satisfies the following algebraic equation,
\begin{equation}\label{eqn:equilibrium}
\b A_i\b x_i^e+\b B_i\left(P_{ci}^{ref} + u_{i}^{e}\right)+\b G_iP_{ei}^e=\b 0,
\end{equation}
where $u_{i}^e=\b k_i(\b x_i^e-\b x_i^d)$. To study the stability of each generator~\eqref{eqn:sysx1} driven by the controller~\eqref{eqn:controller} after disturbances, the perturbed system dynamics about the equilibrium state $\b x_i^e$ is considered. Let $\Delta\b x=[\Delta\b x_1^\top\ \cdots\ \Delta\b x_N^\top]^\top$, where $\Delta\b x_i=\b x_i-\b x_i^e$
is the vector of the deviations of $\delta_i$, $\omega_{ri}$, $P_{mi}$, $X_{mi}$, and $X_{ei}$, respectively, from their equilibrium values after disturbances. Then by subtracting~\eqref{eqn:equilibrium} from~\eqref{eqn:sysx1}, the perturbed dynamics of the $i$-th generator can be derived as
\begin{equation}\label{eqn:sysx2}
\Delta\dot{\b x}_i=\b A_i\Delta\b x_i+\b B_i\Delta u_i+\b G_ih_i(\Delta\b x),
\end{equation}
where $\Delta u_i=u_i-u_i^e=\b k_i\Delta\b x_i$ and $h_i(\Delta\b x)=P_{ei}-P_{ei}^e$. The unknown input $h_i(\Delta\b x)=P_{ei}-P_{ei}^e$ represents external disturbances resulted from network interconnections.

Given the two-axis generator model described in~\cite{Sauer98}, the disturbance $h_i(\Delta\b x)$ can be equivalently represented as
\begin{equation}\label{eqn:hi}
h_i(\Delta\b x)=h_i^{qq}(\Delta\b x)+h_i^{qd}(\Delta\b x)+h_i^{dq}(\Delta\b x)+h_i^{dd}(\Delta\b x)
\end{equation}
with
\begin{align*}
h_i^{qq}(\Delta\b x)&=\sum_{j=1}^NE_{qi}^\prime E_{qj}^\prime\left(G_{ij} \mathcal{C}_{ij} + B_{ij} \mathcal{S}_{ij}\right), \\
h_i^{qd}(\Delta\b x)&=\sum_{j=1}^NE_{qi}^\prime E_{dj}^\prime\left(B_{ij}\mathcal{C}_{ij} - G_{ij}\mathcal{S}_{ij}\right), \\
h_i^{dq}(\Delta\b x)&=\sum_{j=1}^NE_{di}^\prime E_{qj}^\prime\left(-B_{ij}\mathcal{C}_{ij} + G_{ij}\mathcal{S}_{ij}\right), \\
h_i^{dd}(\Delta\b x)&=\sum_{j=1}^NE_{di}^\prime E_{dj}^\prime\left(G_{ij}\mathcal{C}_{ij} + B_{ij}\mathcal{S}_{ij}\right),
\end{align*}
where $E_{qi}^\prime$ is the $q$-axis transient EMF (p.u.), $E_{di}^\prime$ is the $dq$-axis transient EMF (p.u.), $\mathcal{C}_{ij}\triangleq\cos\delta_{ij}-\cos\delta_{ij}^e$, and $\mathcal{S}_{ij}\triangleq\sin\delta_{ij}-\sin\delta_{ij}^e$ with $\delta_{ij}=\delta_i-\delta_j$ and $\delta_{ij}^e=\delta_i^e-\delta_j^e$. Applying the same argument as in~\cite{Siljak02}, it gives
\begin{equation*}\label{eqn:hqqi3}
{h_i^{qq}}^\top(\Delta\b x)h_i^{qq}(\Delta\b x)\le\b y_i^\top\boldsymbol{\mathcal{D}}_i^{qq}\b y_i
\end{equation*}
where $\b y_i=\begin{bmatrix} y_{i1} & \cdots & y_{iN} \end{bmatrix}^\top$ with
\begin{equation*}\label{eqn:yi}
y_{ij}=\frac{\delta_{ij}-\delta_{ij}^e}{2}=\frac{\Delta\delta_i-\Delta\delta_j}{2},
\end{equation*}
and  $\boldsymbol{\mathcal{D}}_i^{qq}=\diag \begin{bmatrix} d_{i1}^{qq} & \cdots & d_{iN}^{qq} \end{bmatrix}$ with
\begin{equation*}\label{eqn:diqq}
d_{ij}^{qq}=4\bar{E}_{qi}^{\prime2}\bar{E}_{qj}^\prime \left(G_{ij}^2+B_{ij}^2\right)^{\frac{1}{2}}
\sum_{k=1}^N\bar{E}_{qk}^\prime\left(G_{ik}^2+B_{ik}^2\right)^{\frac{1}{2}},
\end{equation*}
where $\bar{E}$ represents the allowable maximum absolute value of transient EMF (p.u.). Similarly, the same arguments can be applied, as above, to $h_i^{qd}(\Delta\b x)$, $h_i^{dq}(\Delta\b x)$ and $h_i^{dd}(\Delta\b x)$. Then, it follows from~\eqref{eqn:hi} that
\begin{equation}\label{eqn:hi1}
h_{i}^\top(\Delta\b x)h_{i}(\Delta\b x)\le 4\b y_i^\top\boldsymbol{\mathcal{D}}_i\b y_i,
\end{equation}
where $\boldsymbol{\mathcal{D}}_i=\boldsymbol{\mathcal{D}}_i^{qq}+\boldsymbol{\mathcal{D}}_i^{qd}
+\boldsymbol{\mathcal{D}}_i^{dq}+\boldsymbol{\mathcal{D}}_i^{dd}$. As shown in~\cite{Siljak02}, the inequality~\eqref{eqn:hi1} can be further represented as
\begin{equation*}\label{eqn:hi3}
h_{i}^\top(\Delta\b x)h_{i}(\Delta\b x)\le \Delta\b x^\top\boldsymbol{\mathcal{H}}_i^\top\boldsymbol{\mathcal{H}}_i\Delta\b x,
\end{equation*}
or, more generally,
\begin{equation*}\label{eqn:hi4}
h_{i}^\top(\Delta\b x)h_{i}(\Delta\b x)\le\beta_i^2\Delta\b x^\top\boldsymbol{\mathcal{H}}_i^\top\boldsymbol{\mathcal{H}}_i\Delta\b x,
\end{equation*}
where $\beta_{i}\ge 1$ and $\boldsymbol{\mathcal{H}}_i$ is directly determined from $\boldsymbol{\mathcal{D}}_i$.

Now, the linear feedback gain matrix $\b k_i$ can be obtained by solving a LMI-based optimization problem as formulated in~\cite{Siljak02}. It can be shown using similar arguments as in~\cite{Siljak02} that local decentralized controllers can stabilize the perturbed system~\eqref{eqn:sysx2} if the following optimization problem is feasible,
\begin{equation}\label{eqn:optimization}
\left.\begin{array}{ll}
\textrm{minimize} & \sum_{i=1}^N\left(\gamma_i+\kappa_{Y_i}+\kappa_{L_i}\right) \\
\textrm{subject to} & \b Y_D>0 \\
& \setlength{\arraycolsep}{2pt}\left[\begin{array}{ccccc}
   \b W_D & \b G_D & \b Y_D\b H_1^\top & \cdots & \b Y_D\b H_N^\top \\
   \b G_D^\top & -\b I & \b O & \cdots & \b O \\
   \b H_1\b Y_D & \b O & -\gamma_1\b I & \cdots & \b O \\
   \vdots & \vdots & \vdots & \ddots & \vdots \\
   \b H_N\b Y_D & \b O & \b O & \cdots & -\gamma_N\b I \\
 \end{array}\right]<0 \\
& \left[\begin{array}{cc}
-\kappa_{L_i}\b I & \b L_i^\top \\
\b L_i & -\b I
\end{array}\right]<0, \left[\begin{array}{cc}
\b Y_i & \b I \\
\b I & \kappa_{Y_i}\b I
\end{array}\right]>0 \\
& \gamma_i-\frac{1}{\bar{\beta}_i^2}<0, \quad i=1,\ldots,N,\setlength{\arraycolsep}{5pt}\\
\end{array}\right.
\end{equation}
where $\gamma_i=1/\beta_i$, $\bar{\beta}_i\ge1$, $\b W_D=\b A_D\b Y_D+\b Y_D\b A_D^\top+\b B_D\b L_D+\b L_D^\top\b B_D^\top$ with $\b A_D=\diag \begin{bmatrix} \b A_1 & \cdots & \b A_N \end{bmatrix}$ and $\b B_D=\diag \begin{bmatrix} \b B_1 & \cdots & \b B_N \end{bmatrix}$, and $\kappa_{Y_i}$ and $\kappa_{L_i}$ are prescribed positive limits imposed on the magnitude of $\b L_i$ and $\b Y_i$.

Once the above optimization problem is solved, the gain matrices of local decentralized controllers can be calculated as $\displaystyle \b K_D=\b L_D\b Y_D^{-1}$, where $\b K_D=\diag \begin{bmatrix} \b k_1 & \cdots & \b k_N \end{bmatrix}$. When solving the above optimization problem, the system's tolerance to interconnection uncertainties is maximized.

\subsection{Practical Implementation}

There are two of many possible ways to practically implement the proposed robust controllers for damping the inter-area oscillations. One way is to keep them activated in the system as a pre-emptive strategy to improve the damping ratios of the inter-area oscillation modes. The other way is to install them in the system without activation until poorly damped inter-area oscillations are detected, which can be achieved by the approach of real-time mode estimation as proposed in~\cite{Zhou12}. In this way, the activation of these pre-installed robust controllers becomes part of the remedial action scheme as an emergency measure to prevent potential cascading power failures. The following simulation results confirm the effectiveness under both ways of controller implementation.

\section{Case Studies}\label{sec:case}
In this section, the effectiveness of the proposed decentralized robust control in improving the small signal stability is first demonstrated under both normal and disrupted system operations. In particular, the robustness of the proposed control with respect to the changes in the operating condition and system topology is illustrated through comparative case studies that are simulated by the Power System Toolbox~(PST)~\cite{PSTmanual}. Then the proposed control strategy is also investigated for the case when the power system has only a very limited number of generators equipped with steam valve governors.

\begin{case}\label{case:one}
In this case, the performance of the proposed robust controllers is compared to that of PSSs under different normal operations by varying the operating conditions. The selected test system is an IEEE $2$-area, $4$-machine test system as shown in Fig.~\ref{fig:4machine}, whose prototype was originally introduced in~\cite{Kundur94}.
Two PSSs are installed on generators G$2$ and G$3$, respectively, and a static VAR compensator~(SVC) is connected to bus $101$. The base case has a $400$~MW tie-line flow from the left area to the right area, where the real power of two loads at bus $4$ and bus $14$ is $976$~MW and $1757$~MW, respectively.
\begin{figure}
\centering
\includegraphics[width=0.85\linewidth]{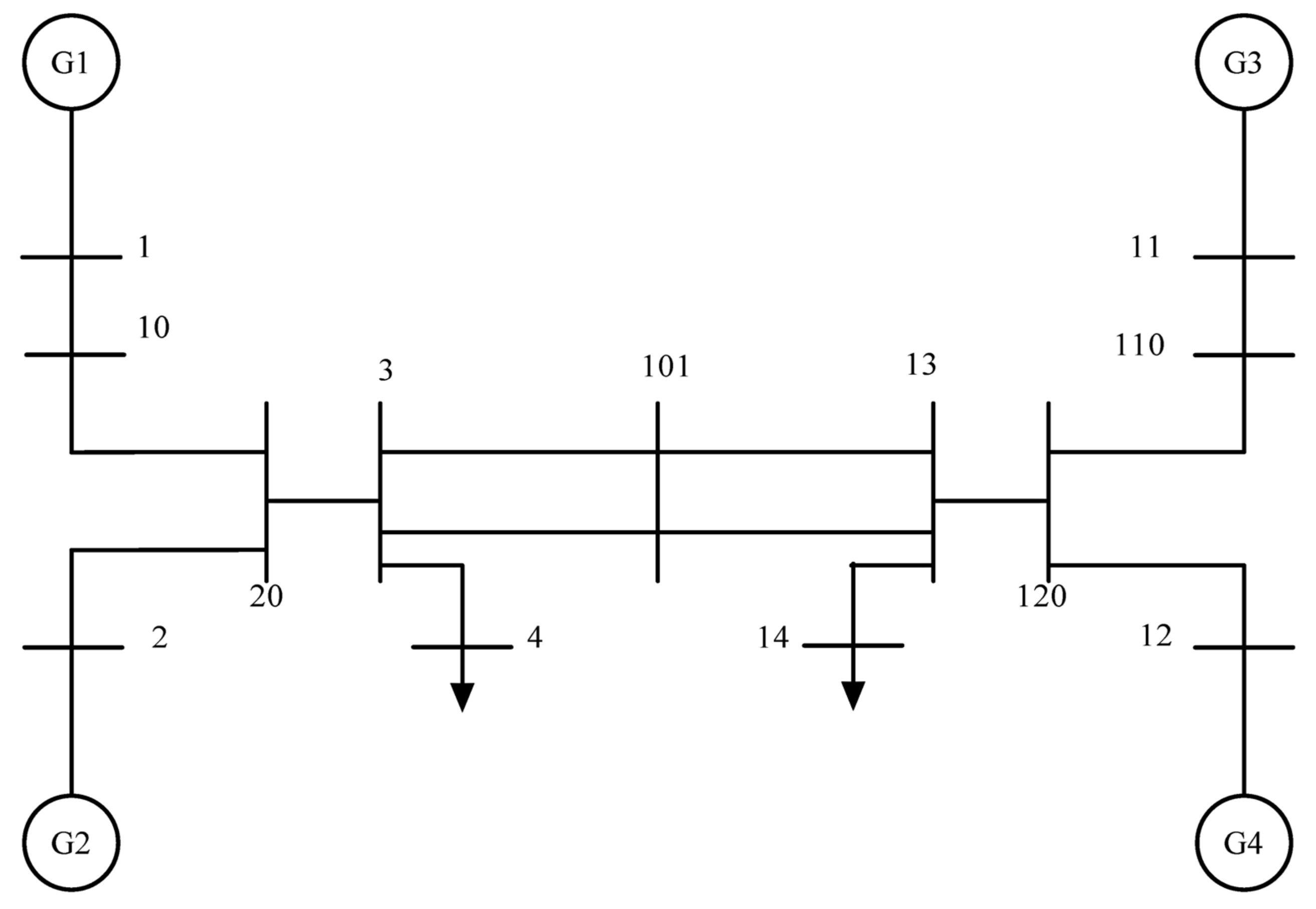}
\caption{Single-line diagram of IEEE $2$-area, $4$-machine test system~\cite{PSTmanual}.}
\label{fig:4machine}
\end{figure}

The modal analysis performed in the PST indicates that the base case with PSSs only has a major inter-area oscillation mode around $0.60$~Hz with a damping ratio of $6.96\%$. This inter-area oscillation mode has the minimum damping ratio among all the oscillation modes of the base case with PSSs only. Since all the generators are equipped with steam valve governors, the proposed decentralized robust controllers can be designed for all the generators. When robust controllers are implemented on generators G$1$ to G$4$, there is an inter-area oscillation mode around $0.70$~Hz with a greatly improved damping ratio of $40.16\%$. However, the oscillation mode of minimum damping ratio is a local mode associated with generators G$1$ and G$2$. When robust controllers are implemented on generators G$1$ and G$4$, there is an inter-area oscillation mode around $0.65$~Hz with a greatly improved damping ratio of $24.10\%$. However, the oscillation mode of minimum damping ratio is again a local mode associated with generators G$1$ and G$2$. When robust controllers are implemented on generators G$2$ and G$3$, the oscillation mode of minimum damping ratio is an inter-area oscillation mode around $0.66$~Hz with a greatly improved damping ratio of $16.09\%$. The minimum damping ratio among all the oscillation modes, which is the most used index for evaluating the small signal stability~\cite{Gomes03}, is listed in Table~\ref{tab:basecase} for each implementation of robust controllers. Therefore, different controller implementations will lead to different improvements in the small signal stability. The greatest improvement of the minimum system damping occurs when robust controllers are implemented on all the generators.
\renewcommand{\arraystretch}{1.2}
\begin{table}
\caption{Oscillation Mode of Minimum Damping Ratio in Base Case for IEEE $2$-area, $4$-machine Test System}
\label{tab:basecase}
\centering
\begin{tabular}{c|c|c|c|c}
\hline
Robust controllers & none & G$1$-G$4$ & G$1$,G$4$ & G$2$,G$3$ \\
\hline
\hline
Oscillation mode & inter-area & local & local & inter-area \\
\hline
Natural frequency & $0.60$ Hz & $1.30$ Hz & $1.20$ Hz & $0.66$ Hz \\
Damping ratio & $6.96\%$ & $23.08\%$ & $18.20\%$ & $16.09\%$ \\
\hline
\end{tabular}
\end{table}\renewcommand{\arraystretch}{1.0}

In order to examine how the minimum damping ratio is correlated with the tie-line flow, various system stress levels are considered by varying the real and reactive power of two loads at buses $4$ and $14$ with the same percentage. At the same time, the power output of generators G$1$ to G$4$ is also changed accordingly by the same percentage. This adjustment pattern creates a number of operating points with different tie-line flows, and it also minimizes the locational effect of generation and load. The corresponding variations of the minimum damping ratio with and without robust controllers are shown in Fig.~\ref{fig:correlation_4}. When the system has PSSs only, the damping ratio of the major inter-area oscillation mode is very sensitive to the system stress level, and decreases significantly with the increase of tie-line flow as shown in Fig.~\ref{fig:correlation_4}(a). However, it can be seen from Fig.~\ref{fig:correlation_4}(b) that the minimum damping ratio of the system is not only greatly improved but also much less sensitive to the system stress level, which leads to a higher power transfer capacity. Thus, the effectiveness of the proposed control in improving the small signal stability is robust to varying operating conditions because no linearization is involved in the proposed controller design.
\begin{figure}
\centering
\subfigure[Without robust controllers]{\includegraphics[width=0.85\linewidth]{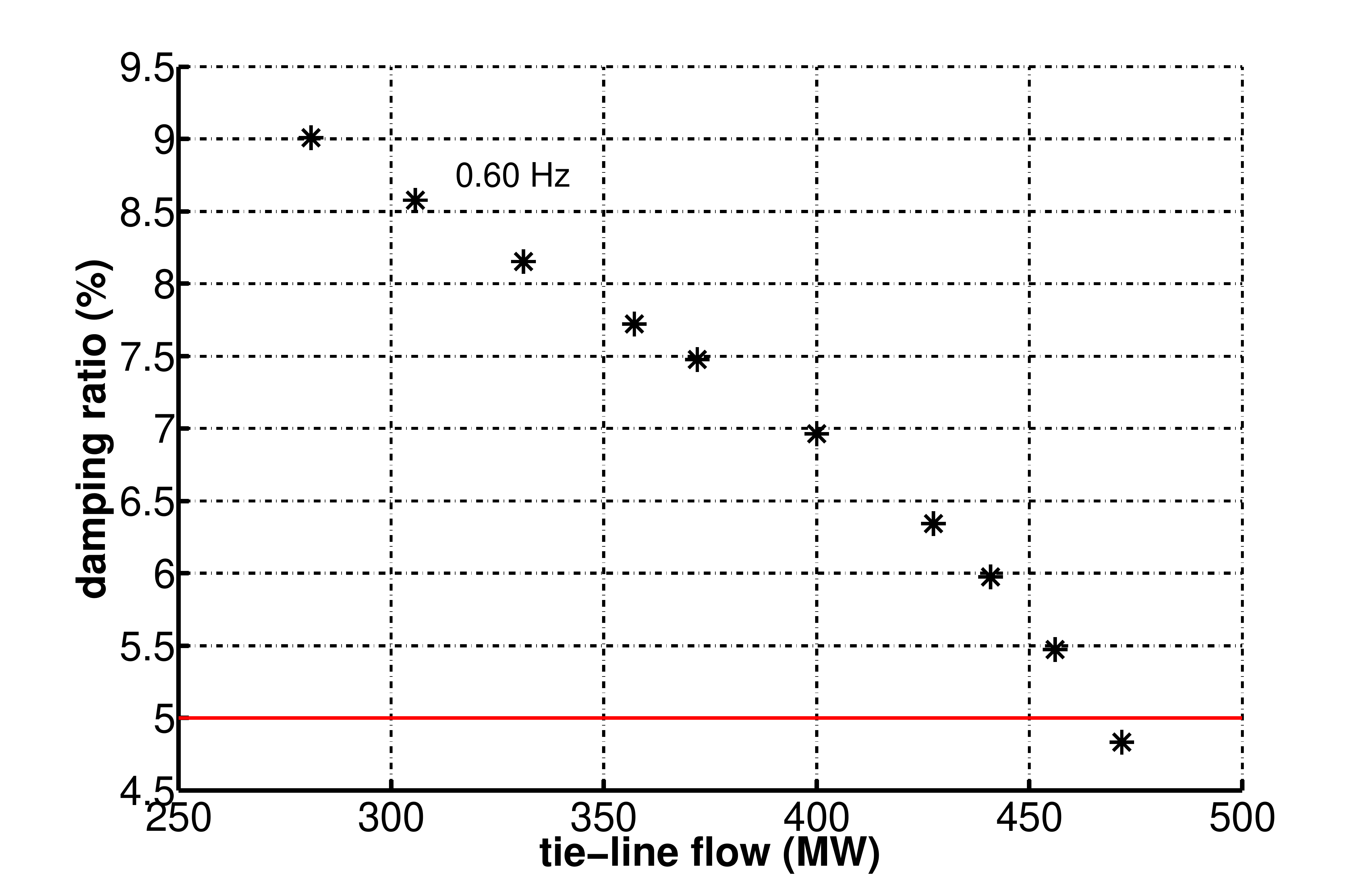}}
\subfigure[With robust controllers]{\includegraphics[width=0.85\linewidth]{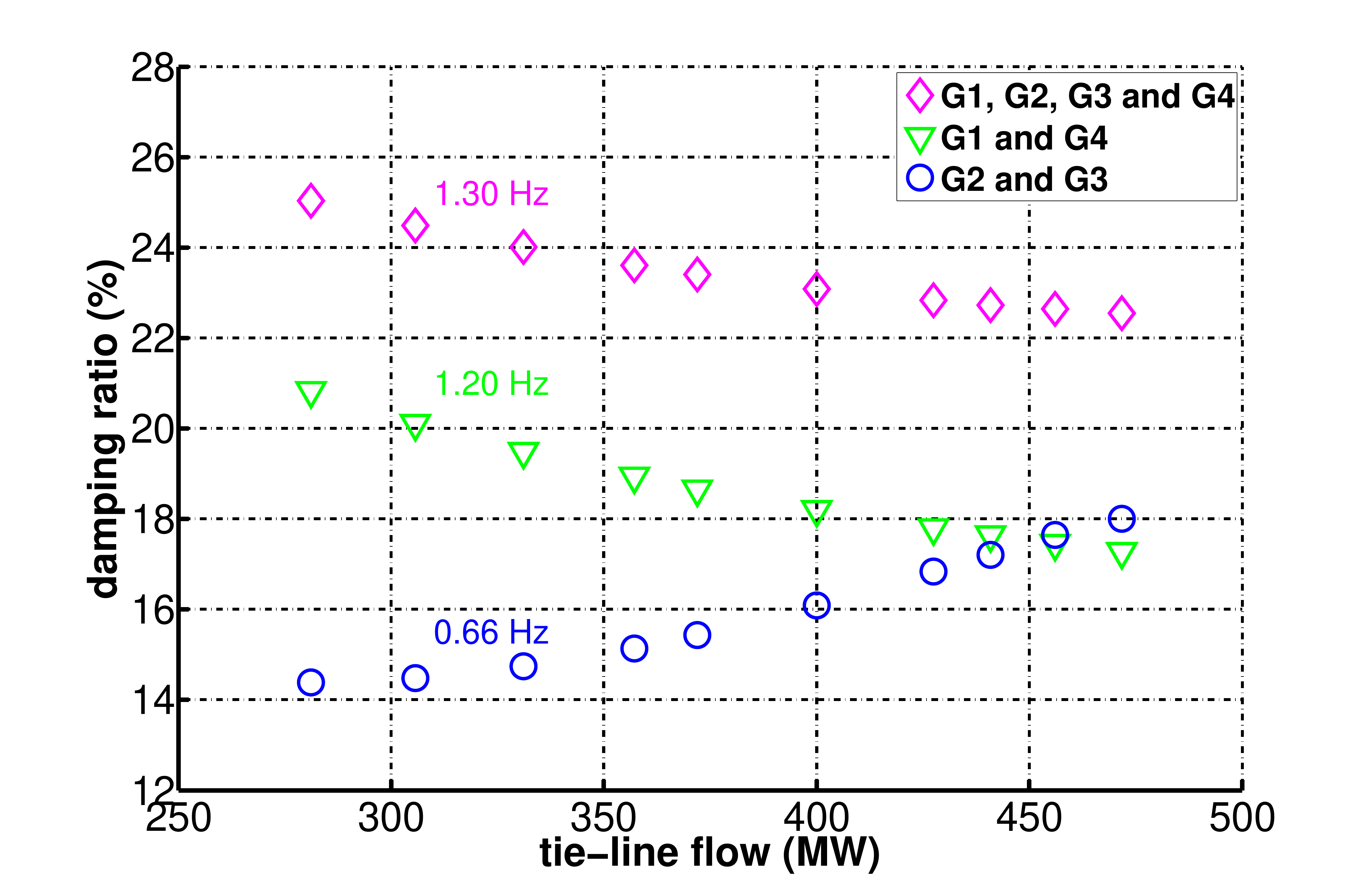}}
\caption{Correlation of the minimum damping ratio with the tie-line flow for the IEEE $2$-area, $4$-machine test system. The red solid line in (a) represents the $5\%$ damping margin.}
\label{fig:correlation_4}
\end{figure}
\end{case}

\begin{case}\label{case:two}
In this case, the performance of the proposed control is compared to that of the PSS under the disrupted system operation by changing the system topology, where the same test system used in Case~\ref{case:one} is considered again. The power demand of two loads and the power output of four generators are increased by $5.58\%$ over the base case. That is, the real power of two loads at bus $4$ and bus $14$ is increased to $1030.4$~MW and $1855$~MW. As perviously determined, this test system with PSSs only has a major inter-area oscillation mode around $0.55$~Hz with a damping ratio of $6.34\%$. Since the damping ratio is larger than the $5\%$ damping margin, any oscillations resulted from small load fluctuations are expected to be well damped. However, the impact of the topology changes due to contingencies on the test system with PSSs only remains to be examined. Thus, in this case, the system topology is changed by tripping a single line between buses $3$ and $101$ without fault. After removing one of the lines $3$-$101$, the test system with PSSs only is identified to have a major oscillation mode around $0.44$~Hz with a damping ratio of $0.46\%$. It turns out that the selected topology change has a very large impact on the test system with PSSs only, which causes a $5.88\%$ damping reduction. Although there are still PSSs on generators G$2$ and G$3$, the resulting system is expected to exhibit a highly oscillatory response, which is dangerous since it may lead to system breakup and large-scale blackouts.

Now consider the test system with robust controllers. When robust controllers are implemented on G$1$ to G$4$, the test system after topology change has a local oscillation mode around $1.30$~Hz with the minimum damping ratio of $23.24\%$. When robust controllers are implemented on generators G$1$ and G$4$, the test system after topology change has a local oscillation mode around $1.18$~Hz with the minimum damping ratio of $18.20\%$. When robust controllers are implemented on generators G$2$ and G$3$, the test system after topology change has an inter-area oscillation mode around $0.48$~Hz with the minimum damping ratio of $16.90\%$. Unlike the PSSs, the proposed control maintains adequate system damping even when the system topology changes.
In order to better illustrate the robustness of the proposed control with respect to the topology changes, the dynamic response of the test system is simulated, respectively, with and without robust controllers.
In the first dynamical simulation, the robust controllers are assumed to be activated in the test system all the time, and the corresponding response of the relative rotor angle between generators G$3$ and G$1$ is shown in Fig.~\ref{fig:dynamics}(a). In the second dynamical simulation, the robust controllers are not online initially. They are only activated after $10$ sec as the reaction to the detection of sustained oscillatory system response. It can be seen from Fig.~\ref{fig:dynamics}(b) that the robust controllers can quickly damp out the oscillations once they are activated.
\begin{figure}
\centering
\subfigure[Robust controllers activated all the time]{\includegraphics[width=0.825\linewidth]{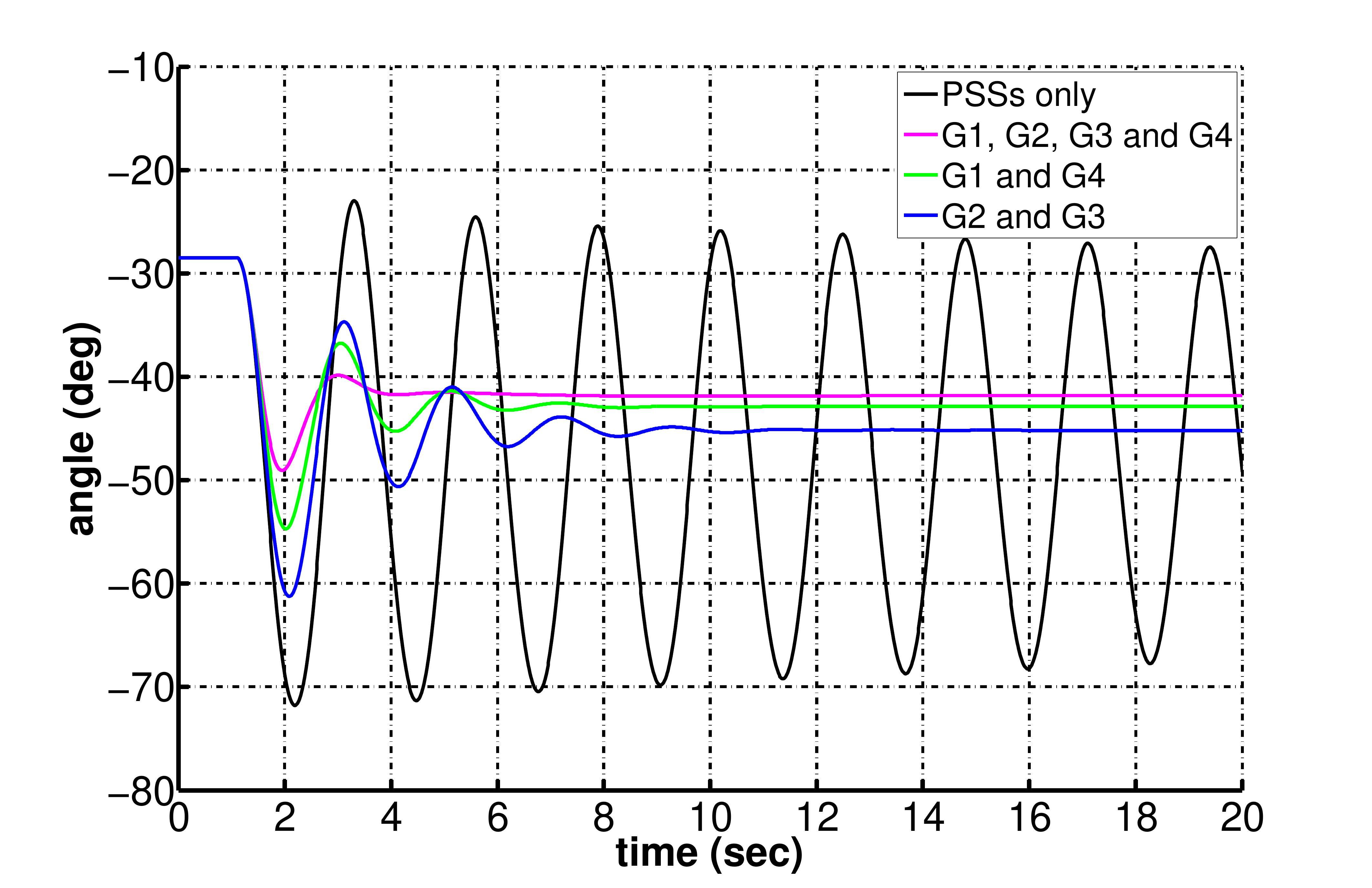}}
\subfigure[Robust controllers activated at $10$ sec]{\includegraphics[width=0.825\linewidth]{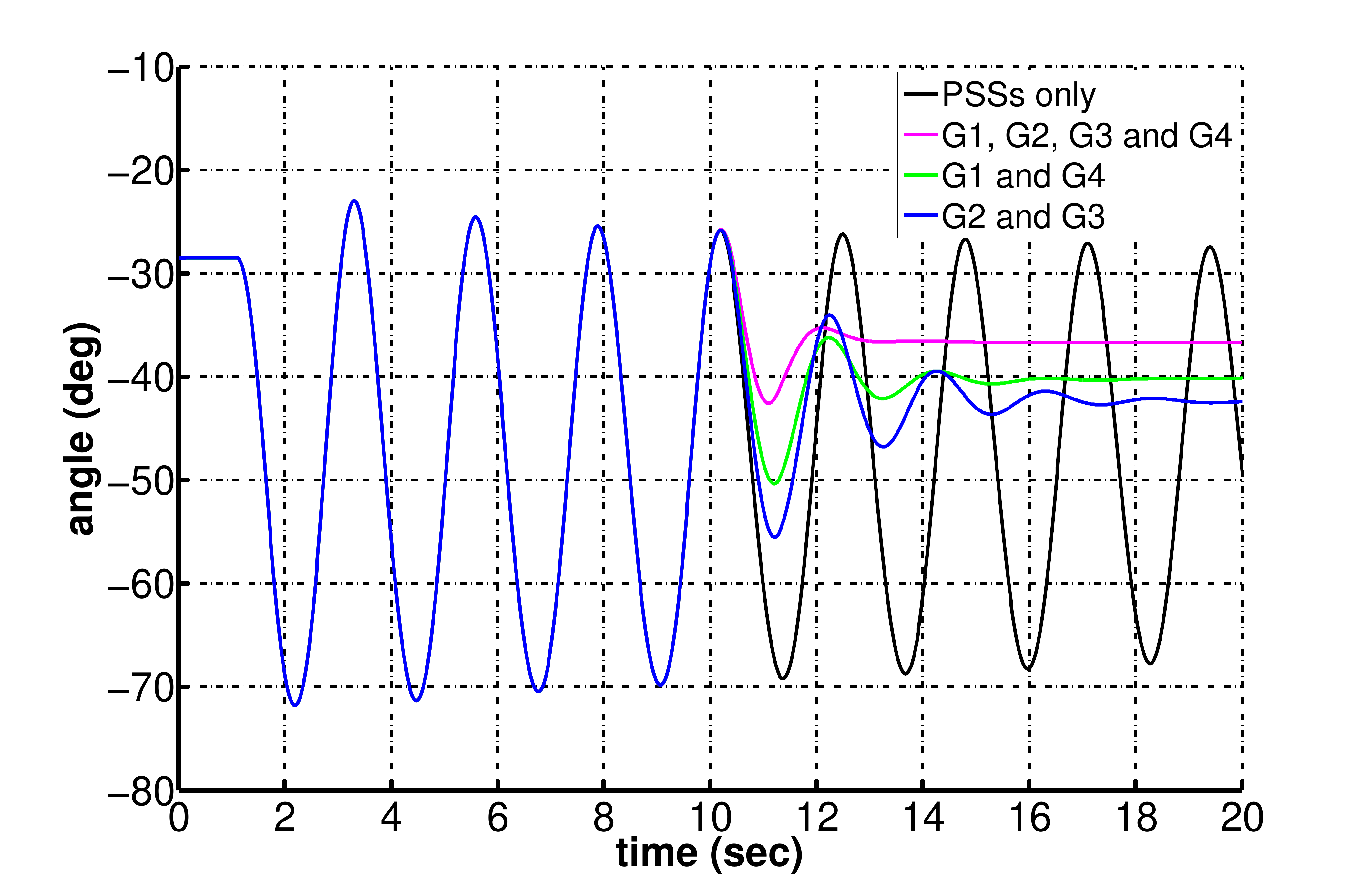}}
\caption{Relative angle $\delta_3$-$\delta_1$ in response to the topology change.}
\label{fig:dynamics}
\end{figure}
\end{case}

\begin{case}\label{case:three}
In this case, the performance of the proposed control is investigated when the power system has only a very small number of generators with steam valve governors. In such a case, the effectiveness of the proposed control could be limited due to the restrictive placement of robust controllers. The selected test system is the $17$-machine system developed in~\cite{Trudnowski08}, which is a modified version of the one originally developed in~\cite{Trudnowski91} as a simplified model of the western North American power grid. This test system as shown in Fig.~\ref{fig:17machine} consists of generator buses $17$ through $24$ and $45$, and load buses $31$ through $41$. All the generators except generator G$17$ are set to work in pairs. Each pair consists of a base generator (G$1$ to G$8$) without speed governor and a load-following generator (G$9$ to G$16$) with speed governor. Only four of the load-following generators (G$11$, G$12$, G$13$ and G$15$) are equipped with steam valve governors, while the rest are equipped with hydraulic governors. Generator G$17$ is a base generator without speed governor. All the generators are equipped with fast voltage regulators and PSSs.
\begin{figure}
\centering
\includegraphics[width=0.825\linewidth]{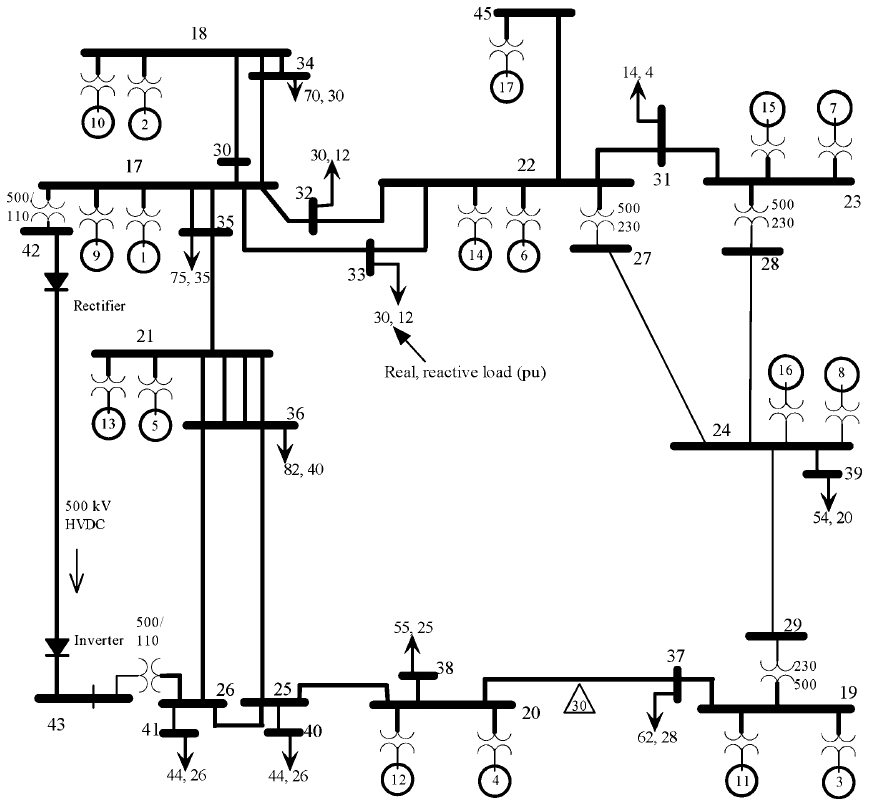}
\caption{Single-line diagram of 17-machine test system~\cite{Trudnowski08}.}
\label{fig:17machine}
\end{figure}

The modal analysis indicates that this test system with PSSs only has four major inter-area oscillation modes around $0.30$~Hz, $0.40$~Hz, $0.60$~Hz and $0.80$~Hz, where the $0.40$~Hz mode has the minimum damping ratio. In order to investigate the correlation between the damping ratios of these inter-area oscillation modes with the system stress level, the power output of generator G$2$ and the power demand at bus $35$ are adjusted by the same percentage, which is $100\%$ for the base case. It can be seen from Fig.~\ref{fig:correlation_17} that there exists a consistent correlation between the damping ratio of the $0.40$~Hz mode and the system stress level when there are only PSSs in the system. Similar correlation also exists for all the other inter-area oscillation modes. When robust controllers are implemented on generators G$11$, G$12$, G$13$ and G$15$, the damping ratios of the $0.30$~Hz, $0.60$~Hz and $0.80$~Hz modes are greatly improved and maintained above the $5\%$ damping margin at various system stress levels. However, it can be seen from Fig.~\ref{fig:correlation_17} that the effectiveness of the proposed control in improving the damping ratio of the $0.40$~Hz mode is not that significant because the required damping margin cannot be achieved when the system is highly stressed.
\begin{figure}
\centering
\includegraphics[width=0.825\linewidth]{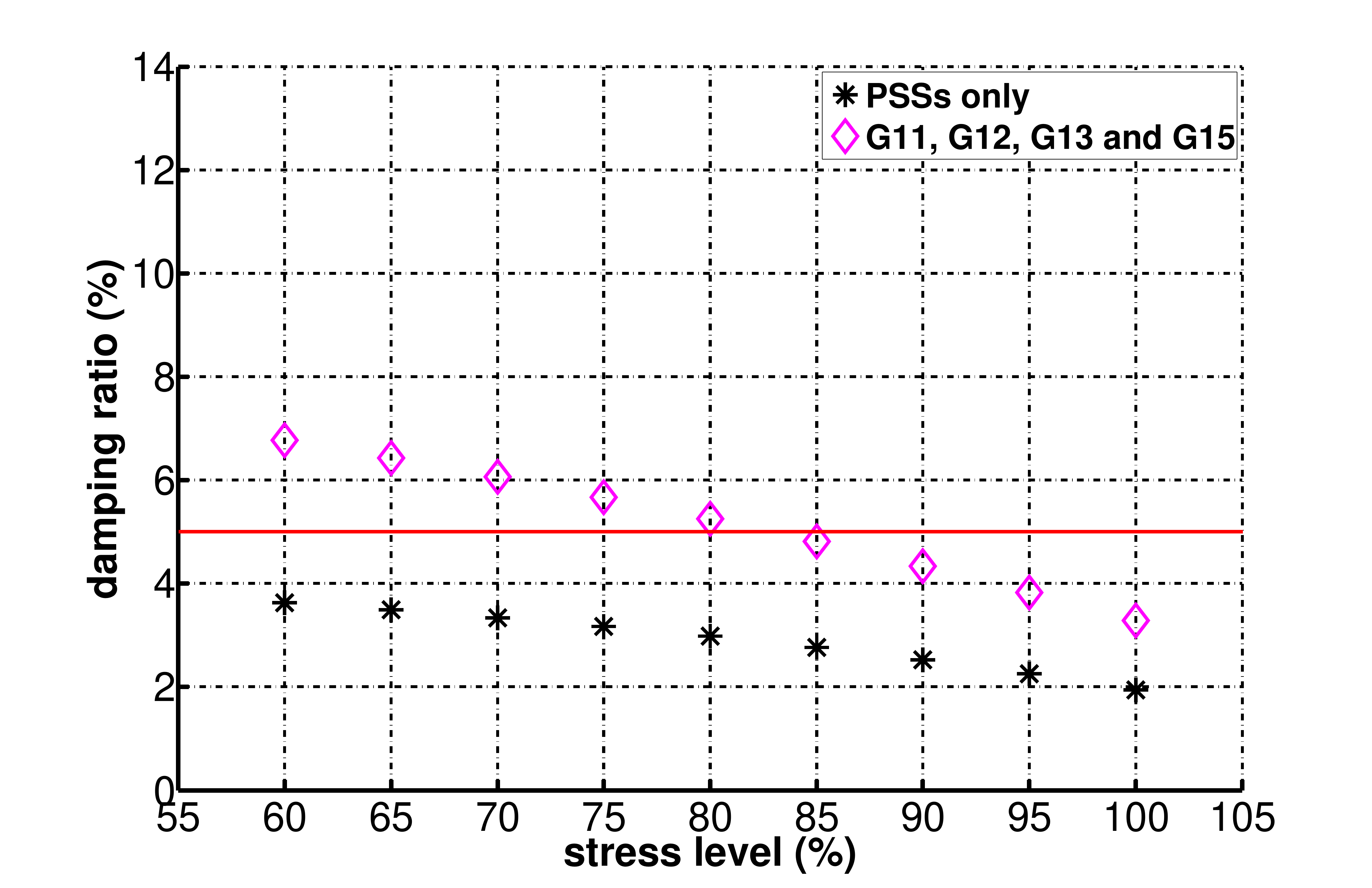}
\caption{Correlation of the damping ratio of the $0.40$~Hz mode with the system stress level for the 17-machine test system.}
\label{fig:correlation_17}
\end{figure}
This is mainly because there are no robust controllers implemented in the areas interconnected by the tie-lines between buses $17$ and $18$. The low damping ratio of the $0.4$~Hz mode is the result of large power flow over the line $17$-$18$. Therefore, the effectiveness of the proposed control requires that robust controllers be implemented in the areas that contribute to the poorly damped inter-area oscillation modes. Since it is not feasible for the given $17$-machine test system to satisfy this requirement, other damping controllers should be deployed in the areas around buses $17$ and $18$ to complement the existing four robust controllers in improving the small signal stability. For example, selective FACTS controllers can be considered to be installed on the line $17$-$18$.

In order to verify the above implementation requirement for the guaranteed effectiveness of the proposed control, the damping ratio of the $0.40$~Hz mode is examined for a modified 17-machine system as shown in Fig.~\ref{fig:18machine}.
\begin{figure}
\centering
\includegraphics[width=0.815\linewidth]{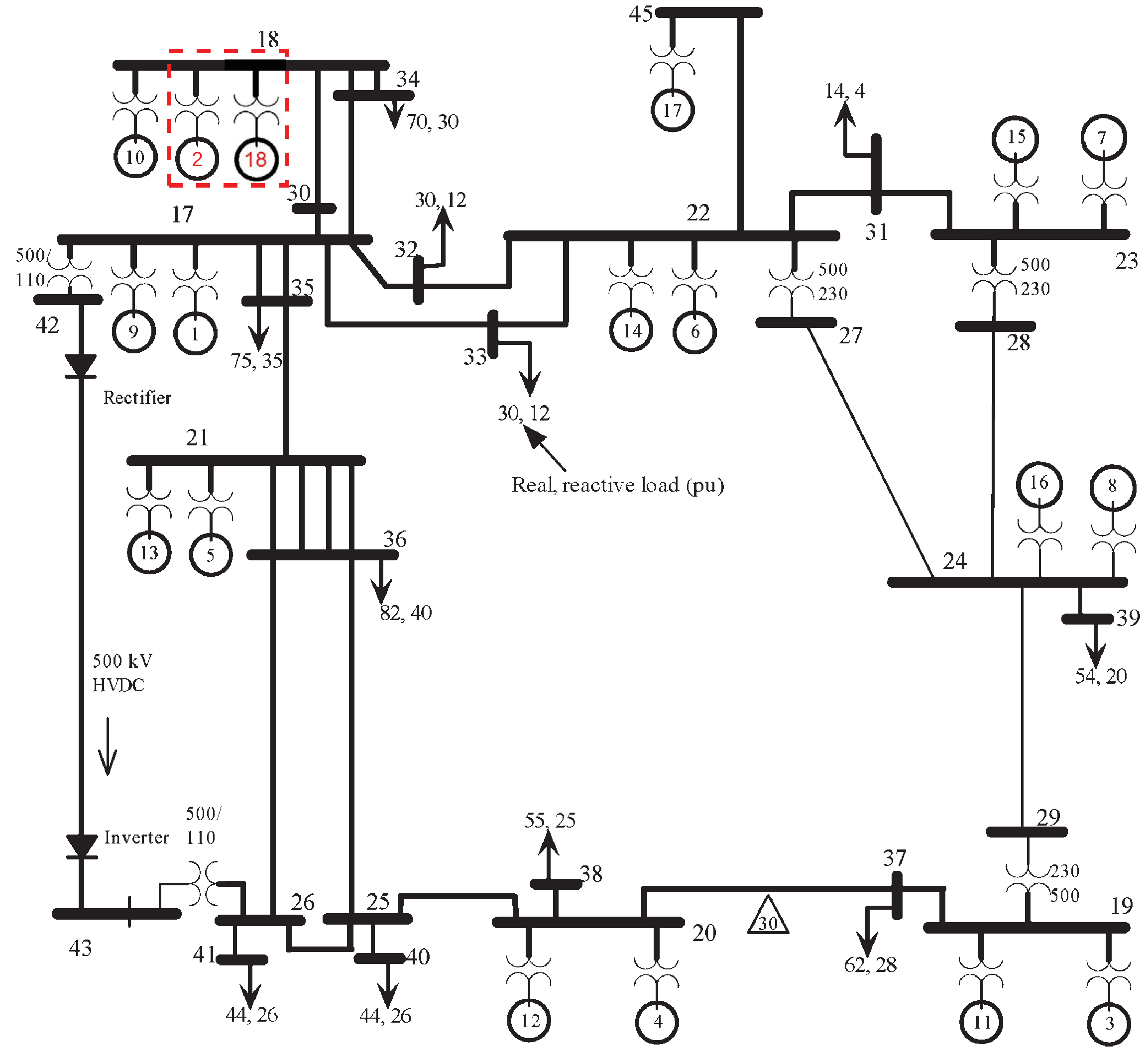}
\caption{Single-line diagram of modified 17-machine test system.}
\label{fig:18machine}
\end{figure}
In this modified system, the original generator G$2$ in Fig.~\ref{fig:17machine} is split into two generators, G$2$ and G$18$. The new G$2$ is still equipped with the hydraulic governor, while the new G$18$ is equipped with the steam valve governor. Their machine parameters are carefully selected such that the damping ratio of the $0.4$~Hz mode for the base case keeps the same after the system modification. As shown in Fig.~\ref{fig:correlation_18}, when an additional robust controller implemented on generator G$18$, the damping ratio of $0.4$~Hz mode is greatly improved and the required damping margin is guaranteed at various system stress levels. It confirms again the robustness of the proposed control to varying operating conditions.
\begin{figure}
\centering
\includegraphics[width=0.825\linewidth]{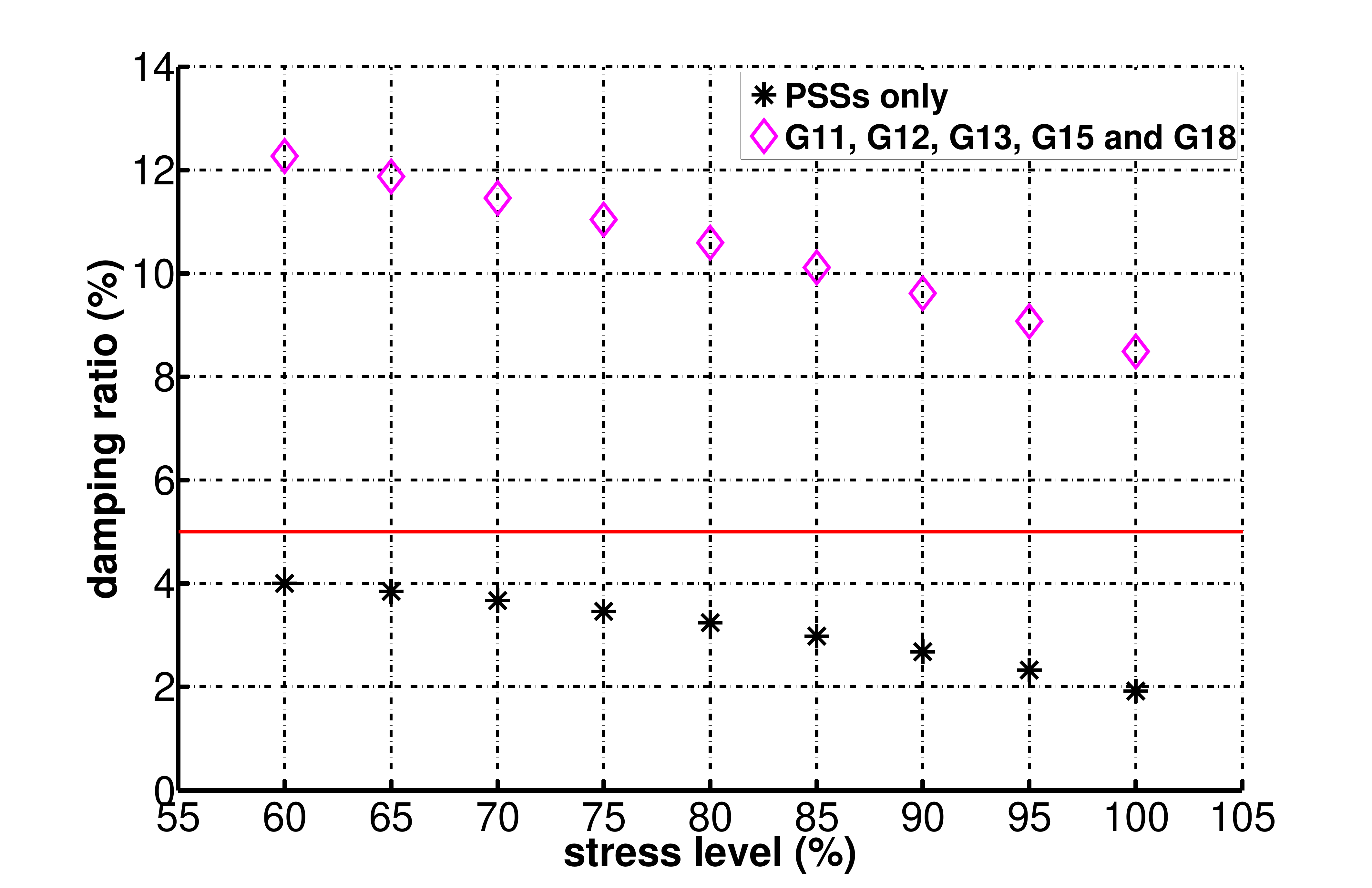}
\caption{Correlation of the damping ratio of the $0.40$~Hz mode with the system stress level for the modified 17-machine test system.}
\label{fig:correlation_18}
\end{figure}

With the modified $17$-machine system in Fig.~\ref{fig:18machine}, the impact of topology changes resulted from $N-1$ contingencies on the damping ratio of the $0.40$~Hz mode for the base case is also evaluated. Total $51$ topology changes are considered with each associated with the tripping of one transmission line. The resulting damping ratio of the $0.40$~Hz mode after contingency is shown in Table~\ref{tab:contigency}, where there are $15$ cases with converged power flow after simple line tripping. It is confirmed again that the effectiveness of the proposed control in improving the small signal stability is robust to the topology change.
\renewcommand{\arraystretch}{1.0}
\begin{table}
\caption{Damping Ratio of the $0.40$~Hz Mode after Simple Line Tripping in Modified 17-machine Test System}
\label{tab:contigency}
\centering
\begin{tabular}{c|c|c}
\hline
Line tripping & PSSs only & G$11$,G$12$,G$13$,G$15$,G$18$ \\
\hline
\hline
$17$-$34$ & $1.10\%$ & $8.21\%$ \\
$18$-$34$ & $0.86\%$ & $7.72\%$ \\
$17$-$32$ & $1.72\%$ & $8.33\%$ \\
$22$-$32$ & $1.85\%$ & $7.99\%$ \\
$17$-$33$ & $1.78\%$ & $8.24\%$ \\
$22$-$33$ & $1.96\%$ & $7.58\%$ \\
$19$-$29$ & $2.73\%$ & $6.56\%$ \\
$25$-$36$ & $3.05\%$ & $8.51\%$ \\
$21$-$36$ & $1.93\%$ & $8.50\%$ \\
$26$-$36$ & $2.71\%$ & $8.47\%$ \\
$22$-$27$ & $1.95\%$ & $8.23\%$ \\
$23$-$28$ & $1.85\%$ & $8.18\%$ \\
$24$-$27$ & $1.95\%$ & $8.23\%$ \\
$24$-$28$ & $1.85\%$ & $8.18\%$ \\
$24$-$29$ & $2.73\%$ & $6.56\%$ \\
\hline
\end{tabular}
\end{table}\renewcommand{\arraystretch}{1.0}
\end{case}

\section{Conclusions}\label{sec:conclusion}
In this paper, decentralized robust controllers have been developed for generators with steam valve governors to improve the damping ratios of the inter-area oscillation modes by directly affecting the real power in the system. The proposed damping control strategy introduces an auxiliary control signal into the governor, creating an additional mechanical torque on the generator rotor. It has several important advantages over the existing damping strategies. The most valuable advantage is the robustness with respect to different operating conditions and system topologies. The controller performance has been demonstrated by detailed case studies on both small and medium-sized test systems. It has been shown that the small signal stability of power systems can be effectively enhanced by the proposed robust control to allow larger power exchange between distinct areas, which in turn leads to higher transmission capacity and better network utilization.

Currently, the proposed control strategy is being extended to include generators with hydraulic governors as well, which are particularly prevalent in the Pacific Northwest. Furthermore, the approach of mitigating the inter-area oscillations directly through real power control is also under investigation for other energy resources such as energy storage or wind generators in addition to conventional generators.



%
\vfill

\end{document}